\documentclass[twocolumn,showpacs,amsfonts,amsmath,amssymb,aps,pra,nofootinbib]{revtex4}

\usepackage{graphicx}
\usepackage{psfrag}

\begin{document}

\title{Two-level atom at finite temperature}

\author{Tomasz Sowi\'nski}
\email{tomsow@cft.edu.pl}
\affiliation{Center for Theoretical Physics of the Polish Academy of Sciences, \\ Al. Lotnik\'ow 32/46, 02-668 Warsaw, Poland}

\date{\today}

\begin{abstract}
Properties of a two-level atom coupled to the quantized electromagnetic field at finite temperature are determined. The analysis is based on a new method (inspired by QED) of describing qubits, developed previously at zero temperature \cite{BiTs}. In this paper, we make a generalization to finite temperature by introducing the Matsubara formalism and the temperature propagators. We analyze the spectral properties of different types of propagators and we derive a direct connection between the temperature propagators and the real time propagators. To show the effectiveness of this method, we calculate the temperature dependence of the polarizability of a two-level atom in the lowest order of perturbation theory and we predict an unexpected sharpening of the resonance. The whole discussion is carried out without the rotating wave approximation.
\end{abstract}
\pacs{11.10.Wx,32.10.Dk,31.15.ap,32.70.Jz}

\maketitle

\section{Introduction}
The main purpose of this paper is to show that a new method of describing qubits within the quantum field theory formalism developed by us in \cite{BiTs} can be also used to describe qubits at finite temperature. The main idea comes from the observation made long time ago by Matsubara \cite{Ma55} that there exists a close analogy between Feynman propagators and the so called temperature propagators. In his paper, Matsubara described general rules of perturbation techniques at non-zero temperature based on Feynman diagrams and used them with great success to describe the system made of electrons and positrons. The Matsubara formalism was further developed by many authors \cite{Ku57,Wa56,Ma59,Lu60} and it has quickly become one of the main tools of modern statistical mechanics \cite{Ab63,Fe71,Ka89,Da97}. In the usual treatment the methods of quantum field theory are exploited to describe systems in which the number of particles is not conserved, i.e. in the grand canonical assemble. In this paper we will adopt this formalism to analyze the temperature properties of one qubit realized by precisely one electron in atomic states. It is not obvious that it is indeed possible. Therefore, at the beginning we present a short introduction to thermal field theory and we give a clear prescription how to proceed.

The influence of thermal fluctuations on the properties of qubits was partially discussed before. First, a very general treatment was based on dissipation phenomena in the damped cavity model \cite{Sa84,Ba86,Sa06}. The problem of the influence of the temperature on the induced interactions between qubits via the electromagnetic field and on their entanglement is recently considered in the context of the so called {\it sudden death of entanglement''} phenomenon \cite{Al07,La07,Qa08}. Very often this analysis is carried out in the Jaynes–-Cummings model in the rotating wave approximation (RWA) \cite{Bos01,Ki02,Li03,Rei08}. There exists also some numerical predictions on the behavior of qubits at finite temperatures realized in quantum dots \cite{Li08}. So far a general problem of the dynamical properties of a single qubit interacting with the quantized electromagnetic field (without RWA) at finite temperature has not been discussed since all current techniques did not give such possibilities. In this paper we want to show that methods ingrained in quantum field theory could be the first step to solve this problem. 

To make all our argumentation clear, we restrict ourselves to the problem of a two-level atom and its polarizability (response tensor), since a two-level atom is the simplest model of the interaction between light and matter where purely mathematical problems are as small as possible \cite{BiTs}. Moreover, the properties of the linear polarizability of a two-level atom were discussed in detail at zero-temperature regime by many authors and many discussions on it took place \cite{An98,Bu00,St01,Bu01,An03,Mi04,Ber06,Mi08}. This discussions have never went beyond zero temperature case, since it seemed to be a very hard task. In this paper we want to show that our new method of describing qubits in terms of field operators is a key to manage this problem. To do it correctly we will not make RWA at any point of the analysis \cite{Mi04}.

\section{Physical situation}
\subsection{The Hamiltonian of the system}
The starting point of our analysis is the well known Hamiltonian of the two-level atom interacting with the quantized electromagnetic field
\begin{equation}  \label{HamiltonianStandard}
\widehat{\cal H} = \widehat{\cal H}_0 + \widehat{\cal H}_\mathtt{I}
\end{equation}
where
\begin{subequations}
\begin{align}
\widehat{\cal H}_0 &= m\sigma_z +\int_0^\infty\!\!\mathrm{d}k\, k\, a^\dagger(k)a(k),   \label{FreeHamiltonianStandard} \\
\widehat{\cal H}_\mathtt{I} &=\sigma_x\,\int_0^\infty\!\!\mathrm{d}k\, g(k)\,\Phi(k). \label{IntHamiltonianStandard}
\end{align}
\end{subequations}
The Hamiltonian $\widehat{\cal H}_0$ describes the properties of noninteracting subsystems -- two-level atom with the energy gap equal $2m$ and the free quantized electromagnetic field. The operators $a^\dagger(k)$ and $a(k)$ create and annihilate photons in the appropriate modes of the electromagnetic field and together they form a quantum scalar field
\begin{equation} \label{FieldOperatorPhoton}
\Phi(k) = \frac{a(k)+a^\dagger(k)}{\sqrt{2k}}.
\end{equation}

The Hamiltonian $\widehat{\cal H}_\mathtt{I}$ is responsible for the interaction between the photons and the two-level atom. The formfactor $g(k)$ is simply a coupling coefficient measuring the strength of a dipole transition in the atom. Note that we do not make a rotating wave approximation in the coupling Hamiltonian \eqref{IntHamiltonianStandard}. As was shown previously in \cite{Mi04} the complete interaction term is necessary to obtain a correct expression for the polarizability of a two-level atom.

\subsection{Two-level atom in thermal equilibrium}
In our treatment of a qubit at non-zero temperature we will make two physical assumptions. First, that our system described by the Hamiltonian \eqref{HamiltonianStandard} is in thermal equilibrium with the heat reservoir (thermostat) at temperature $T$. Second, that the temperature is not so high as to induce the creation of real electron-positron pairs. It means that the number of fermions in our system is conserved. Notice however, that we do not assume that the number of photons in the system is conserved.

Description of the statistical properties of a quantum system in thermodynamic equilibrium is realized in terms of the statistical assembles. In our model an appropriate assemble is defined by the following control parameters:
\begin{itemize}
\item the number of fermions (electrons) equals $1$,
\item the temperature $T$ of the thermostat,
\item the chemical potential $\mu$ of the photon reservoir.
\end{itemize}
It is worth to notice that because of the properties of the electromagnetic field, the chemical potential $\mu$ is always equal to $0$. It means that all statistical properties of the photons can be expressed as functions of the temperature $T$ and one extensive parameter (for example the volume of the cavity). Nevertheless, to emphasize the fact that $\mu$ is one of our controll parameter we will write it explicitly to the end of this section.

\subsection{Statistical operator}
From the rules of quantum statistical physics (cf., for example, \cite{Hu78}) we know  that any system in equilibrium with the thermostat and the reservoir of particles is in the quantum state represented by the following density matrix (in this context also called statistical operator)
\begin{subequations} \label{Stat1}
\begin{equation} \label{Stat1a}
\widehat\rho = \frac{1}{\widehat{\cal Z}}\mathrm{e}^{-\beta \widehat{\cal K}}, \qquad \widehat{\cal Z}=\mathrm{tr}\left[\mathrm{e}^{-\beta \widehat{\cal K}}\right].
\end{equation}
As always $\beta=1/kT$ and $\widehat{\cal Z}$ is partition function. In our case the statistical Hamiltonian $\widehat{\cal K}$ is given by
\begin{equation} \label{Stat1b}
\widehat{\cal K} = \widehat{\cal H} - \mu {\cal N},
\end{equation}
\end{subequations}
where the operator $\cal N$ represents the number of photons in the system. Expectation value of any operator $\cal O$ in such a quantum state of the system is given by
\begin{align} \label{Stat2}
\langle\!\langle {\cal O}\rangle\!\rangle = \mathrm{tr}\big[ \widehat\rho\,{\cal O}\big] = \frac{\mathrm{tr}\big[ \mathrm{e}^{-\beta\widehat{\cal K}} {\cal O}\big]}{\mathrm{tr}\big[\mathrm{e}^{-\beta\widehat{\cal K}}\big]}.
\end{align}

For future needs let us now introduce an additional system composed of a noninteracting two-level atom and the electromagnetic field. We define the free statistical Hamiltonian $\widehat{\cal K}_0$ and the free statistical operator $\widehat{\rho}_0$ of such a system as follows
\begin{subequations} \label{Stat9}
\begin{align}
\widehat{\cal K}_0 &= \widehat{\cal H}_0 - \mu {\cal N}, \\
\widehat{\rho}_0 &= \frac{1}{\widehat{\cal Z}_0}\,\mathrm{e}^{-\beta \widehat{\cal K}_0}, \qquad \widehat{\cal Z}_0=\mathrm{tr}\left[\mathrm{e}^{-\beta \widehat{\cal K}_0}\right].
\end{align}
\end{subequations}
The statistical operator $\widehat{\rho}_0$ describes a quantum state of a system composed of a free two-level atom and  the free electromagnetic field which together are in equilibrium with the thermostat and the reservoir of photons. The connection between statistical Hamiltonians in the interacting and the noninteracting case is following
\begin{equation}
\widehat{\cal K} = \widehat{\cal K}_0 + \widehat{\cal H}_\mathtt{I}
\end{equation}
and the expectation value of the operator $\cal O$ in the state $\widehat{\rho}_0$ is given by
\begin{equation}
\langle\!\langle {\cal O}\rangle\!\rangle_0 = \mathrm{tr}\big[ \widehat{\rho}_0\,{\cal O}\big] = \frac{\mathrm{tr}\big[ \mathrm{e}^{-\beta\widehat{\cal K}_0} {\cal O}\big]}{\mathrm{tr}\big[\mathrm{e}^{-\beta\widehat{\cal K}_0}\big]}.
\end{equation}

\subsection{Linear polarizability of the atom}
The linear polarizability of a two-level atom is a physical measure of the reaction of the system to small external electromagnetic perturbations. When the state of the system is described by the statistical operator \eqref{Stat1a}, then the polarizability is defined as the following two-point retarded correlation function \cite{Zu60,Lo06}
\begin{equation} \label{AlphaTTReal}
\mathfrak{a}(t,t') = -iA\,\theta(t-t')\,\left\langle\!\!\left\langle \left[\mathrm{e}^{i\widehat{\cal H}t}\sigma_x\,\mathrm{e}^{-i\widehat{\cal H}t},\mathrm{e}^{i\widehat{\cal H}t'}\sigma_x\,\mathrm{e}^{-i\widehat{\cal H}t'}\right]\right\rangle\!\!\right\rangle.
\end{equation}
For a single two-level atom, the constant $A$ is equal to $d^2/2\hbar$ where $d$ is a dipole-moment of the atomic transition. It is easy to find that $\mathfrak{a}(t,t')$ depends only on the difference of its arguments and therefore it has a very simple form in the frequency domain
\begin{equation} \label{AlphaTT}
\mathfrak{a}(\omega) = \int_{-\infty}^\infty \mathrm{d}t\,\mathrm{e}^{i\omega(t-t')}\,\mathfrak{a}(t,t').
\end{equation}
In this paper we give a precise and clear prescription how to find this quantity as a perturbation series using the well known methods of thermal quantum field theory \cite{Ma55,Fe71,Ka89,Ab63,Da97}.

\section{Second quantization}
The basic ingredients of quantum field theory are obviously the field operators. Therefore, before we start to analyze our system in this language, we have to introduce, besides the electromagnetic field operator \eqref{FieldOperatorPhoton}, the fermionic field operator representing the states of an electron in the atom. As usual, field operators can be introduced after performing second-quantization of electronic states. In this model this procedure is very easy. We just enlarge the two dimensional Hilbert space spanned by vectors $|\mathtt{g}\rangle$ and $|\mathtt{e}\rangle$ representing ground state and exited state of the electron in the atom to the four dimensional one spanned by four state vectors: two previously defined $|\mathtt{g}\rangle$ and $|\mathtt{e}\rangle$, the zero-electron state $|\mathtt{N}\rangle$, and the antisymmetric two-electron state $|\mathtt{B}\rangle$ representing two electrons -- one in each state. 

Such a procedure enables one to introduce creation and annihilation operators of the electron. The operators $\psi_\downarrow$ and $\psi^\dagger_\downarrow$ annihilate and create electron in the ground state of the two-level atom, and $\psi_\uparrow$ and $\psi^\dagger _\uparrow$ do so in the exited state. These operators form together the fermion field operator and its hermitian conjugate \cite{BiTs}
\begin{equation} \label{FieldOperatorFermion}
  \Psi =\left(\begin{array}{c}\psi_\uparrow \\ \psi_\downarrow \end{array}\right), \qquad \Psi^\dagger =\left(\psi^\dagger_\uparrow \, , \, \psi^\dagger_\downarrow \right).
\end{equation}
Then one can easily find the second-quantized form of the new Hamiltonian acting in the extended Hilbert space:
\begin{subequations} \label{Hamiltonian}
\begin{equation} 
{\cal H} = {\cal H}_0 + {\cal H}_\mathtt{I}
\end{equation}
where
\begin{align}
{\cal H}_0 &= m \Psi^\dagger \sigma_z \Psi+\int_0^\infty\!\!\mathrm{d}k\, k\, a^\dagger(k)a(k),   \label{FreeHamiltonian} \\
{\cal H}_\mathtt{I} &=\Psi^\dagger\sigma_x\,\Psi\int_0^\infty\!\!\mathrm{d}k\, g(k)\,\Phi(k). \label{IntHamiltonian}
\end{align}
\end{subequations}
The origin, and the properties of the Hamiltonian \eqref{Hamiltonian} and of the field operators $\Psi$ and $\Psi^\dagger$ are discussed in detail in our previous paper \cite{BiTs}. Notice that the system described by the Hamiltonian $\cal H$ may have states with different number of fermions. However, the number of fermions operator commutes with $\cal H$.

In the Heisenberg picture, the time evolution of the system is contained in the dynamics of the field operators. They evolve according to the following equations \cite{BiTs}:
\begin{subequations}
\begin{align}
\left( i\partial_t -m\sigma_z\right)\Psi(t) &= \int_0^\infty\!\!\mathrm{d}k\,g(k)\, \Phi(k,t)\sigma_x\Psi(t), \label{HPsiDyn}\\
\left(\partial_t^2+k^2\right)\Phi(k,t) &= -g(k)\,\Psi^\dagger(t)\sigma_x\Psi(t). \label{HPhiDyn}
\end{align}
\end{subequations}

Now we introduce a new statistical operator $\rho$ representing the thermal equilibrium of the new quantum mechanical system
\begin{subequations}
\begin{equation}
\rho = \frac{1}{\cal Z}\mathrm{e}^{-\beta \cal K}, \qquad {\cal Z}=\mathrm{tr}\left[\mathrm{e}^{-\beta \cal K}\right]
\end{equation}
where the statistical Hamiltonian $\cal K$ is given by
\begin{equation}
{\cal K} = {\cal H} - \mu {\cal N}.
\end{equation}
\end{subequations}
The expectation value of any operator $\cal O$ in the quantum state $\rho$ is given by
\begin{equation} \label{Stat2X}
\langle {\cal O} \rangle = \mathrm{tr}\big[ \rho\,{\cal O}\big] = \frac{\mathrm{tr}\big[ \mathrm{e}^{-\beta{\cal K}} {\cal O}\big]}{\mathrm{tr}\big[\mathrm{e}^{-\beta{\cal K}}\big]}.
\end{equation}

Notice, that the expectation values \eqref{Stat2} and \eqref{Stat2X}, even for the same physical observable ${\cal O}$, are different since the statistical Hamiltonians $\cal K$ and $\widehat{\cal K}$ act in different Hilbert spaces. Nevertheless, for those qubit operators ${\cal O}_1,\ldots,{\cal O}_n$ that are represented by traceless $2 \times 2$ matrices (for example the free qubit Hamiltonian $m_0\sigma_z$, the interaction Hamiltonian \eqref{IntHamiltonianStandard}, or any other linear combination of the Pauli matrices) one can verify that there exists the following direct connection between the expectation value of their product $\langle\!\langle {\cal O}_1\cdots{\cal O}_n\rangle\!\rangle$ and expectation value of their second-quantized counterparts $\langle \Psi^\dagger{\cal O}_1\Psi\cdots\Psi^\dagger{\cal O}_n\Psi\rangle$
\begin{equation} \label{correspondence}
\langle\!\langle {\cal O}_1\cdots{\cal O}_n \rangle\!\rangle = \frac{\cal Z}{\widehat{\cal  Z}}\,\langle \Psi^\dagger{\cal O}_1\Psi\cdots\Psi^\dagger{\cal O}_n\Psi \rangle.
\end{equation}
It should be emphasized that this relation is a central theorem of this paper. It expresses nontrivial but a very simple connection between the expectation values of physical quantities of the system containing exactly one fermion and the expectation values of their  non-physical second-quantized counterpart. Therefore, it is a peculiar link between the interesting physical system described by the Hamiltonian \eqref{HamiltonianStandard} and a convenient mathematical model described by the Hamiltonian \eqref{Hamiltonian}. Without this connection further analysis would not be possible.

Relation \eqref{correspondence} follows from the observation that for traceless matrices the bilinear combination of field operators $\Psi^\dagger {\cal O}_i \Psi$ gives zero when acting on the two-electron or the zero-electron states. Moreover, in the qubit (one-electron) subspace it acts as the qubit operator ${\cal O}_i$. Obviously, this argument is also valid for evolving qubit operators ${\cal O}_i(t)=\mathrm{e}^{i\widehat{\cal H}t}{\cal O}_i\mathrm{e}^{-i\widehat{\cal H}t}$ and their counterparts $\Psi^\dagger(t)\,{\cal O}_i\,\Psi(t)$.

This all means that a linear polarizability of the two-level atom defined by traceless $\sigma_x$ matrices in \eqref{AlphaTTReal} can be reproduced from its second-quantized counterpart
\begin{equation} \label{AlphaTTT}
\alpha(t,t') = -iA\,\theta(t-t')\,\left\langle \left[\Psi^\dagger(t)\sigma_x\,\Psi(t),\Psi^\dagger(t')\sigma_x\,\Psi(t')\right]\right\rangle
\end{equation}
by using the property \eqref{correspondence}. 

Note that function $\alpha(t,t')$ has no direct physical sense, since the statistical operator $\rho$ involves also non physical states. Therefore we stress that we treat $\alpha(t,t')$ only as a very convenient tool to find a physical quantity.

\section{Matsubara formalism}
The starting point of the formulation of quantum field theory at finite temperatures is to introduce the temperature Green functions (or temperature correlation functions). They are defined in an analogous way as the Feynman Green functions at zero temperature. The difference lies in the evolution parameter -- instead of the ordinary time variable $t$ we use the imaginary time $\tau=i t$. As was first noticed by Matsubara in 1955 \cite{Ma55} after this simple replacing one can use almost all known machinery of the quantum field theory. In analogous to the ordinary time case all analysis is carried out in two quantum mechanical pictures -- Heisenberg-like picture and Dirac-like picture.

In the Matsubara-Heisenberg picture the whole evolution of the system is contained in the evolution of the operators. Any operator ${\cal O}$ evolves according to the following  rule
\begin{equation} \label{FT5A}
{\cal O}(\tau) = \mathrm{e}^{{\cal K} \tau}\,{\cal O}\,\mathrm{e}^{-{\cal K} \tau}.
\end{equation}
In particular, the field operators \eqref{FieldOperatorPhoton} and \eqref{FieldOperatorFermion} evolve in the following way
\begin{subequations} \label{FT5}
\begin{align}
\Phi(k,\tau) &= \mathrm{e}^{{\cal K} \tau}\,\Phi(k)\,\mathrm{e}^{-{\cal K} \tau}, \\
\Psi(\tau) &= \mathrm{e}^{{\cal K} \tau}\,\Psi\,\mathrm{e}^{-{\cal K} \tau}, \\
\Psi^\dagger(\tau) &= \mathrm{e}^{{\cal K} \tau}\,\Psi^\dagger\,\mathrm{e}^{-{\cal K} \tau}. \end{align}
\end{subequations}

In contrast, in the Matsubara-Dirac picture the operators evolve according to the free statistical Hamiltonian ${\cal K}_0$
\begin{equation} \label{FT6}
O(\tau) = \mathrm{e}^{{\cal K}_0 \tau}\,{\cal O}\,\mathrm{e}^{-{\cal K}_0 \tau}.
\end{equation}
One can easily check that the field operators \eqref{FieldOperatorPhoton} and \eqref{FieldOperatorFermion} evolve in the following way (from now we will put $\mu=0$ everywhere)
\begin{subequations} \label{FT6b}
\begin{align}
\psi(\tau) &=\left(\begin{array}{c}\psi_\uparrow\,\mathrm{e}^{-m\tau} \\ \psi_\downarrow\,\mathrm{e}^{m\tau} \end{array}\right), \\[5pt]
\psi^\dagger(\tau)&=\left(\psi_\uparrow^\dagger\, \mathrm{e}^{m\tau},\psi_\downarrow^\dagger\,\mathrm{e}^{-m\tau}\right),  \\[2pt]
\mathrm{\phi}(k,\tau) &= \frac{a(k)\mathrm{e}^{-k \tau}+a^\dagger(k)\mathrm{e}^{k \tau}}{\sqrt{2k}}. \label{FT6bc}
\end{align}
\end{subequations}
The interaction Hamiltonian \eqref{IntHamiltonian} in this picture has the form
\begin{align}
{\cal H}_{\mathtt{I}}(\tau) &= \mathrm{e}^{{\cal K}_0 \tau}\,{\cal H}_{\mathtt{I}}\,\mathrm{e}^{-{\cal K}_0 \tau} \nonumber \\
&= \psi^\dagger(\tau)\,\sigma_x\,\psi(\tau) \int_0^\infty\!\!\mathrm{d}k\,\, g(k)\mathrm{\phi}(k,\tau).
\end{align}

\subsection{Temperature correlation functions} \label{TempGreenFunct}
Temperature correlation functions are the central objects of the quantum field theory at finite temperatures. In analogy to the Feynman ones, they do not have any direct physical interpretation, but they are the basic ingredients of the perturbation computations based on Feynman diagrams. 

We call the $n$-point temperature correlation function a complex function of $n$ parameters defined as follows (To underline the difference between imaginary and ordinary time functions we use a ,,tilde''): 
\begin{equation} \label{FT7}
\widetilde{\cal G}(\tau_1,\ldots,\tau_n) = -\langle \mathbb{T}_\tau\, {\cal O}(\tau_1)\cdots{\cal O}(\tau_n)\rangle.
\end{equation}
where the operator $\mathbb{T}_\tau$ is the time-ordering operator. This operator respects the fermionic nature of operators, and therefore for any odd permutation of electron operators one should remember about changing the sign of the expression. It will be clarified later that in practice it is sufficient to restrict each imaginary-time variable $\tau_i$ only to the range
\begin{equation} \label{taurestrictions}
0\leq\tau_i\leq\beta.
\end{equation}

Special case of temperature correlation functions \eqref{FT7} are temperature propagators of electromagnetic and fermion fields. They are defined as follows
\begin{subequations} \label{FT8}
\begin{align}
\widetilde{\cal S}_{\alpha\beta}(\tau_1-\tau_2) &=\widetilde{\cal S}_{\alpha\beta}(\tau_1,\tau_2) \nonumber \\
&= -\langle \mathbb{T}_\tau\, \Psi_\alpha(\tau_1)\Psi^\dagger_\beta(\tau_2)\rangle, \label{FT8a}\\
\widetilde{\cal D}(k,k',\tau_1-\tau_2)&=\widetilde{\cal D}(k,k',\tau_1,\tau_2) \nonumber \\
&= -\langle \mathbb{T}_\tau\, \Phi(k,\tau_1)\Phi(k',\tau_2)\rangle. \label{FT8b}
\end{align}
\end{subequations}

The temperature propagators are functions of the difference of their arguments only. Therefore, according to the restriction \eqref{taurestrictions} they are defined on the finite range of length $2\beta$. Moreover, they fulfill the symmetric boundary condition $\widetilde{\cal G}(-\beta)=\widetilde{\cal G}(\beta)$. Therefore one can represent them as the following Fourier series \cite{Ab63,Fe71,Ka89,Da97}:
\begin{subequations} \label{FT10}
\begin{equation} \label{FT10a}
\widetilde{\cal G}(\tau) = \frac{1}{\beta} \sum_{n=-\infty}^\infty \mathrm{e}^{-i\omega_n \tau}\, \widetilde{\cal G}(\omega_n),
\end{equation}
where
\begin{equation} \label{FT10b}
\omega_n = \left\{
\begin{array}{cl}
\frac{2n\pi}{\beta} & \textrm{bosons}, \\[10pt]
\frac{2(n+1)\pi}{\beta} & \textrm{fermions}.
\end{array}
\right.
\end{equation}
\end{subequations}
The Fourier coefficients $\widetilde{\cal G}(\omega_n)$ can be easily reconstructed from the inverse Fourier transform
\begin{equation} \label{FT12}
\widetilde{\cal G}(\omega_n) = \int_0^\beta\!\mathrm{d}\tau\,\mathrm{e}^{i\omega_n\tau}\widetilde{\cal G}(\tau).
\end{equation}

Since all properties of a two-level atom will be determined via the free temperature propagators we will now calculate them in detail. Similarly as it was done at zero temperature \cite{BiTs}, free propagators are defined in terms of the field operators in the Matsubara-Dirac picture:
\begin{subequations} \label{TempFree}
\begin{align}
\widetilde{S}_{\alpha\beta}(\tau_1,\tau_2) &= -\langle \mathbb{T}_\tau\, \psi_\alpha(\tau_1)\psi^\dagger_\beta(\tau_2)\rangle_0, \label{TempFreeElektron}\\
\widetilde{D}(k,k',\tau_1,\tau_2) &= -\langle \mathbb{T}_\tau\, \mathrm{\phi}(k,\tau_1)\mathrm{\phi}(k',\tau_2)\rangle_0. \label{TempFreeFoton}
\end{align}
\end{subequations}

To find the free temperature electron propagator one has to use the averages of the bilinear combinations of the fermionic field operators over the statistical assemble. Then by simple calculations one finds the following representation of the electron propagator:
\begin{equation}
\widetilde{S}(\tau_1,\tau_2)  =-\left(\frac{\theta(\tau_1-\tau_2)}{1+\mathrm{e}^{-m\beta\sigma_z}}-\frac{\theta(\tau_1-\tau_2)}{1+\mathrm{e}^{m\beta\sigma_z}}\right)\mathrm{e}^{-(m\sigma_z)(\tau_1-\tau_2)}.
\end{equation}
Consequently the Fourier coefficient of this propagator reads
\begin{equation}
\widetilde{S}(\omega_n) = \frac{1}{i\omega_n -m\sigma_z}, \qquad \mathrm{where} \qquad \omega_n = \frac{(2n+1)\pi}{\beta}.
\end{equation}

Similarly, by using appropriate averages of bosonic field operators, one can get the free propagator of the electromagnetic field. It is not hard task to find the free temperature photon propagator and to show that its Fourier component reads
\begin{align}
\widetilde{D}(k,k',\omega_n) =-\frac{\delta(k-k')}{\omega_n^2+k^2},\qquad \mathrm{where} \qquad \omega_n = \frac{2n\pi}{\beta}.
\end{align}

\subsection{Perturbation series and Feynman rules}
The methods of quantum field theory at finite temperature are based on theorems which are formally very similar to the standard theorems used at zero temperature. This observation gives us a possibility to adapt automatically all well known machinery to this more complicated physical situation.

The main relation which gives us a chance to start perturbation calculations is a generalized Gell-Mann and Low theorem. It gives us explicitly a connection between the temperature correlation functions in the Matsubara-Heisenberg picture and the corresponding correlation functions in the Matsubara-Dirac picture. This fundamental relation is \cite{Ab63,Fe71}
\begin{multline} \label{GellMann}
\langle \mathbb{T}_\tau {\cal O}_1(\tau_1)\cdots{\cal O}_n(\tau_n)\rangle \\= \frac{
\langle \mathbb{T}_\tau O_1(\tau_1)\cdots O_n(\tau_n)\,\mathrm{e}^{-\int_0^\beta\!\mathrm{d}\tau\, {\cal H}_\mathtt{I}(\tau)}\rangle_0}
{\langle \mathbb{T}_\tau \,\mathrm{e}^{-\int_0^\beta\!\mathrm{d}\tau\, {\cal H}_\mathtt{I}(\tau)}\rangle_0}.
\end{multline}
One can easily check that the formal structure of this relation corresponds to the structure of the corresponding equation at zero temperature \cite{BiTs}. Since the integration on the right hand side is done only over the range $\left[0,\beta\right]$, now it is clear why we restricted in \eqref{taurestrictions} the imaginary-time parameters to this range. Formal expansion of the right hand side in the powers of interaction Hamiltonian leads us directly to the concept of Feynman diagrams which are just simple graphical representations of the appropriate elements of the perturbation series. As usual any temperature correlation function in the Matsubara-Heisenberg picture can be expressed as a infinite sum of connected Feynman diagrams.

The finite-temperature Feynman rules for studied qubit system are very simple. At first one should draw Feynman diagrams according to the same rules as were used at zero temperature \cite{BiTs}. Then one assigns them corresponding mathematical expressions. It can be done in the frequency domain by applying following rules:
\begin{itemize}
\item Each photon line between any two vertices represents the free temperature photon propagator
$$
-\widetilde{D}(k,k',\omega_n)=\frac{\delta(k-k')}{\omega_n^2+k^2}, 
$$
where $\omega_n=\frac{2n \pi}{\beta}$.
\item Each fermion line between any two vertices represents the free temperature fermion propagator
$$
-\widetilde{S}(\omega_n)=\frac{-1}{i\omega_n-m\sigma_z},
$$
where $\omega_n=\frac{(2n+1) \pi}{\beta}$.
\item Each vertex represents the expression
$$
-V(k) = - g(k)\sigma_x.
$$
\item In each vertex the frequency conservation law holds. It is a consequence of the imaginary time translation invariance of the system. This rule is a counterpart of the energy conservation law at zero temperature.
\item The fermion propagators are $2\times 2$ matrices and therefore they should be multiplied in the order given by the directions of the lines.
\item Each closed fermion loop leads to the trace operation of the matrices and gives a factor $-1$. This rule is a simple consequence of the fermionic nature of the electron filed operators.
\end{itemize}
In the end one should integrate such an expression over all internal momenta $k$ and take a sum over all internal frequencies $\omega_n$. Moreover each sum must be divided by $\beta$.

\section{Temperature photon propagator}
Now we will find the temperature photon propagator in the lowest order of perturbation theory. Since there exists a formal connection between all Feynman rules at finite and zero temperatures, we can automatically write down the fundamental relation between photon propagator in the Matsubara-Heisenberg and in the Matsubara-Dirac pictures. This relation reads \cite{BiTs}
\begin{subequations} \label{FT20}
\begin{multline}
\widetilde{\cal D}(k,k',\omega_n) = \widetilde{D}(k,k',\omega_n)+\\ \int_0^\infty\!\!\mathrm{d}k_1\!\!\int_0^\infty\!\!\mathrm{d}k_2\,\widetilde{D}(k,k',\omega_n)\widetilde{\Pi}(k_1,k_2,\omega_n)\widetilde{\cal D}(k_2,k',\omega_n).
\end{multline}
 
We introduced here the so called temperature self-energy function $\widetilde{\Pi}$ which is built from all irreducible connected Feynman diagrams with two external photon lines
\begin{multline}
-\widetilde{\Pi}(k_1,k_2,\omega_n) = \includegraphics[scale=0.7]{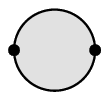} =\,\includegraphics[scale=0.7]{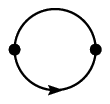}+\,\includegraphics[scale=0.7]{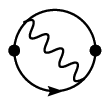}\,+\,\includegraphics[scale=0.7]{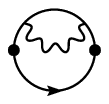}\,+\,\ldots
\end{multline}
\end{subequations}
One can transform the relation \eqref{FT20} to a more convenient form
\begin{subequations}
\begin{align}
\widetilde{\cal D}(k,k',\omega_n) &= -\frac{\delta(k-k')}{\omega_n^2+k^2} \nonumber \\
&+\frac{g(k)}{\omega_n^2+k^2}\left[\frac{1}{\widetilde{\mathrm{P}}(\omega_n)^{-1}-\widetilde{h}(\omega_n)}\right]\frac{g(k')}{\omega_n^2+k'^2},
\end{align}
where we introduced a new function
\begin{equation}
\widetilde{h}(\omega_n) = - \int_0^\infty\!\!\mathrm{d}k\, \frac{g^2(k)}{\omega_n^2+k^2}.
\end{equation}
\end{subequations}
We also define the temperature transition matrix $\widetilde{\mathrm{T}}$
\begin{equation} \label{TPConect}
-\,\widetilde{\mathrm{T}}(\omega_n) = \frac{1}{\includegraphics[scale=0.8]{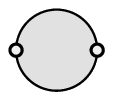}^{-1} -\includegraphics[scale=0.8]{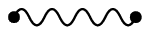} } = -\,\frac{1}{\widetilde{\mathrm{P}}^{-1}(\omega_n)-\widetilde{h}(\omega_n)}.
\end{equation}

Now we are ready to find the lowest order correction to the temperature Feynman photon propagator. The temperature self-energy function of the photon in this order is represented by the following Feynman diagram
\begin{align}
-\widetilde{\mathrm{P}}^{(2)}(\omega_n) &=  \includegraphics[scale=0.8]{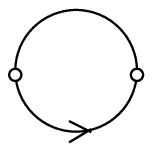} \nonumber \\[10pt] &=-\frac{1}{\beta}\sum_{n'} \mathrm{Tr}\big\{
\sigma_x \widetilde{S}(\omega_{n'}+\omega_n)\sigma_x \widetilde{S}(\omega_{n'})\big\}.
\end{align}
The summation oven $n'$ can be easily done and the details of these calculations are presented in Appendix B. We rewrite a result \eqref{TLA-P-2TApp} derived there as follows
\begin{align} \label{TLA-P-2T}
\widetilde{\mathrm{P}}^{(2)}(\omega_n)&=- \frac{4m}{4m^2+\omega_n^2}\tanh\left(\frac{\beta m}{2}\right).
\end{align}
From the eq. \eqref{TPConect} one can easily find the temperature transition matrix
\begin{equation} \label{TLA-T-2T}
\widetilde{\mathrm{T}}^{(2)}(\omega_n) = -\frac{4m}{(4m^2+\omega_n^2)\coth\left(\frac{\beta m}{2}\right)-4m\widetilde{h}(\omega_n)}.
\end{equation}
It is worth to notice that taking the zero-temperature limit ($\beta\rightarrow \infty$) at this point of our analysis has no physical sense. This fact will be explained later.

\section{Real time propagators}
Let us now return to the problem of the linear polarizability of the atom and also introduce the so called retarded photon propagator. For the operators in the Heisenberg picture this propagator is defined as follows:
\begin{subequations} \label{RetPhotProp}
\begin{equation}
{\cal D}_\mathtt{R}(k,k',t,t') = -i \theta(t-t')\langle \left[ \Phi(t),\Phi(t')\right]\rangle
\end{equation}
and for the free photon field (in the interaction picture) it reads
\begin{equation} \label{RetFreeFoton}
D_\mathtt{R}(k,k',t,t') = -i \theta(t-t')\langle \left[ \phi(t),\phi(t')\right]\rangle_0.
\end{equation}
\end{subequations}

There exists a general connection between the retarded photon propagator defined above and the polarizability of the atom given by \eqref{AlphaTTT}. This connection follows from the equation of motion \eqref{HPhiDyn}, a definition of the polarizability \eqref{AlphaTTT}, and the properties of the free retarded photon propagator \eqref{RetFreeFoton}. The leading argument is identical as at zero temperature and it was given by us in \cite{BiTs}. Therefore we will not repeat it here. This connection has a form
\begin{widetext}
\begin{equation} \label{PropagatorAlphaConnection}
{\cal D}_\mathtt{R}(k,k',k_0) = {D}_\mathtt{R}(k,k',k_0)
-\frac{1}{A}\int_0^\infty\!\!\mathrm{d}k_1\, g(k_1)\!\int_0^\infty\!\!\mathrm{d}k_2\, g(k_2)\,D_\mathtt{R}(k,k_1,k_0)\alpha(k_0)D_\mathtt{R}(k_2,k',k_0)
\end{equation}
\end{widetext}
where ${\cal D}_\mathtt{R}(k,k',k_0)$ and $D_\mathtt{R}(k,k',k_0)$ are the Fourier transforms of the retarded photon propagators defined by eq. \eqref{RetPhotProp}
\begin{subequations} \label{PropRetT}
\begin{align}
{\cal D}_\mathtt{R}(k,k',k_0) &= \int_{-\infty}^{\infty}\!\!\mathrm{d}t\, \mathrm{e}^{ik_0t}\,{\cal D}_\mathtt{R}(k,k',t,0), \label{PropRetTa}\\
D_\mathtt{R}(k,k',k_0)
&= \int_{-\infty}^{\infty}\!\!\mathrm{d}t\, \mathrm{e}^{ik_0t}\,D_\mathtt{R}(k,k',t,0). \label{PropRetTb}
\end{align}
\end{subequations}
From the eq. \eqref{PropagatorAlphaConnection} it clearly follows that we need an appropriate retarded photon propagator to find the polarizability of a two-level atom (or more generally -- the linear response of the system).

\subsection{Spectral representation of propagators}
Now we will demonstrate that there exists an exact mathematical connection between the real time and the temperature propagators. For future benefits we restrict ourself just to the retarded propagator of the electromagnetic field. Other propagators, like the advanced propagator or the chronologicaly ordered propagator, can be also reconstructed from the temperature propagators \cite{Ab63,Fe71,Ka89,Da97} but in the case of the linear polarizability of the atom they are not interesting. 

From the equation \eqref{FT12} one can find Fourier transform of the temperature photon propagator defined by \eqref{FT8b}
\begin{widetext}
\begin{align}
\widetilde{\cal D}(k,k',\omega_n) &= \int_0^\beta \mathrm{d}\tau \mathrm{e}^{i\omega_n\tau}\widetilde{\cal D}(k,k',\tau,0) \nonumber \\
&=\int_0^\beta \mathrm{d}\tau \mathrm{e}^{i\omega_n\tau}\frac{1}{\cal Z}\sum_{n,m}\langle n|\mathrm{e}^{-\beta{\cal K}}\mathrm{e}^{{\cal K}\tau}\Phi(k)\mathrm{e}^{-{\cal K}\tau}|m\rangle\langle m|\Phi(k')|n\rangle \nonumber \\
&=\sum_{n,m}\int_0^\beta \mathrm{d}\tau \mathrm{e}^{(i\omega_n+K_n-K_m)\tau}\frac{\mathrm{e}^{-\beta K_n}}{\cal Z}\langle n|\Phi(k)|m\rangle\langle m|\Phi(k')|n\rangle.
\end{align}
\end{widetext}
In this sequence of equalities we have used the set $\left\{|n\rangle\right\}$ of the eigenvectors of the statistical Hamiltonian ${\cal K}$ with the eigenvalues $K_n$. After performing an integration over $\tau$, one finds
\begin{equation} \label{ProTempSpektr}
\widetilde{\cal D}(k,k',\omega_n)= \int_{-\infty}^\infty\!\!\mathrm{d}M\, \frac{\mathfrak{M}(M,k,k')}{i\omega_n - M},
\end{equation}
where the temperature spectral matrix of the propagator is given by
\begin{multline} \label{SpectralMatrix}
\mathfrak{M}(M,k,k') \\ = \sum_{n,m} \delta(M+K_n-K_m) \frac{\mathrm{e}^{-\beta K_m}-\mathrm{e}^{-\beta K_n}}{\cal Z}\\ \times\langle n|\Phi(k)|m\rangle\langle m|\Phi(k')|n\rangle.
\end{multline}

In an analogous way one can show that retarded propagator \eqref{PropRetTa} has the following spectral representation
\begin{align}
{\cal D}_\mathtt{R}(k,k',t,0) &=-i\theta(t)\Big[\langle \Phi(k,t)\Phi(k')\rangle-\langle \Phi(k')\Phi(k,t)\rangle\Big] \nonumber \\
&=\int_{-\infty}^{\infty} \frac{\mathrm{d}k_0}{2\pi}\mathrm{e}^{-ik_0t}\int_{-\infty}^\infty \!\!\mathrm{d}M\,\frac{\mathfrak{M}(M,k,k')}{k_0-M+i\epsilon}.
\end{align}
It means that the Fourier transform of the retarded photon propagator reads
\begin{equation} \label{ProRetTSpektr}
{\cal D}_\mathtt{R}(k,k',k_0)=\int_{-\infty}^\infty \!\!\mathrm{d}M\,\frac{\mathfrak{M}(M,k,k')}{k_0-M+i\epsilon}.
\end{equation}
\subsection{The analytic continuation}
The spectral representation of the propagators is very useful to find the connection between different types of propagators in the frequency domain. First notice that the retarded propagator \eqref{ProRetTSpektr} treated as a function of the continuous complex variable $k_0$ is analytic in the upper half of the complex plane.  It means that if one defines the function $F(z)$ of the complex variable $z$,
\begin{equation}
F(z) = \int_{-\infty}^\infty \!\!\mathrm{d}M\,\frac{\mathfrak{M}(M,k,k')}{z-M},
\end{equation}
then this function in the upper halfplane  would be exactly equal to the retarded photon propagator \eqref{ProRetTSpektr}. Additionally, at points $z_n=i\omega_n$ this function is equal to the Fourier components of the temperature photon propagator \eqref{ProTempSpektr}. Therefore the temperature photon propagator for positive $\omega_n$ can be easily reconstructed directly from the retarded propagator in the following way
\begin{equation}
\widetilde{\cal D}(k,k',\omega_n)={\cal D}_\mathtt{R}(k,k',i\omega_n).
\end{equation}
The fourier components for the negative frequencies can be reconstructed as well because the  following relation holds
\begin{equation} \label{CrossRelation}
\widetilde{\cal D}(k,k',-\omega_n) =\widetilde{\cal D}^*(k,k',\omega_n).
\end{equation}

The inverse problem, how to construct the retarded propagator from the temperature propagator, is more complicated. The reason is that the Fourier components of the temperature propagator are defined only on a discrete set of frequencies $\omega_n$. Therefore, there exist many analytical continuations of the temperature propagator to the whole complex plane and one needs to add other conditions to choose an appropriate continuation. This general problem was discussed in the past \cite{Ba61,Ab63,Fe71,Da97} and now it is known how to manage it. In our case the retarded propagator can be constructed with the following rule
\begin{equation}
{\cal D}_\mathtt{R}(k,k',k_0) = F(-ik_0+\epsilon),\qquad \textrm{for}\qquad k_0>0.
\end{equation}
To find the retarded propagator for the negative $k_0$ one has to use the crossing relation given in our previous paper \cite{BiTs}.

\section{Polarizability of the two-level atom}
In the previous section we have found a connection between the temperature photon propagator and the retarded photon propagator. It was very important step because it gives us a possibility to find the temperature properties of the linear polarizability of the atom. To make it obvious let us recall once more a fundamental formulas for different types of the photon propagators in the theory with interactions
\begin{widetext}
\begin{subequations} \label{PropTZest}
\begin{align}
\widetilde{\cal D}(k,k',\omega_n) = &\,\widetilde{D}(k,k',\omega_n)+ \frac{g(k)}{k^2+\omega_n^2}\,\widetilde{\mathrm{T}}(\omega_n)\,\frac{g(k')}{k^2+\omega_n^2}, \label{PropTZestA}\\
{\cal D}_\mathtt{R}(k,k',k_0) = &\,D_\mathtt{R}(k,k',k_0) -\frac{1}{A} \frac{g(k)}{k^2-(k_0-i\epsilon)^2}\,\alpha(k_0)\,\frac{g(k')}{k^2-(k_0-i\epsilon)^2}. \label{PropTZestB}
\end{align}
\end{subequations}
\end{widetext}
For positive $k_0$ we can reconstruct the retarded propagator \eqref{PropTZestB} from the temperature one \eqref{PropTZestA} by making the substitution $\omega_n\rightarrow -ik_0+\epsilon$. Therefore, it is self-evident that the linear polarizabilty $\alpha(k_0)$ can be extracted from the temperature transition matrix $\widetilde{\mathrm{T}}(k_0)$. The rule is the following:
\begin{equation}
\alpha(k_0) = - A\, \widetilde{\mathrm{T}}(-ik_0+\epsilon), \qquad \textrm{for}\qquad k>0.
\end{equation}
Once we have determined the temperature transition matrix in the second order of perturbation \eqref{TLA-T-2T}, we are able to find the polarizability in this order. For positive $\omega$ one gets
\begin{equation*}
\alpha^{(2)}(\omega) = \frac{4mA}{(4m^2-\omega^2)\coth\left(\frac{\beta m}{2}\right)-4m\left[\Delta(\omega)+i\Gamma(\omega)\right]}.
\end{equation*}
Functions $\Delta(\omega)$ and $\Gamma(\omega)$ are defined as real and imaginary part of the function $h(\omega)=\widetilde{h}(-i\omega+\epsilon)$ 
\begin{subequations}\label{DeltaIGamma}
\begin{align}
\Delta(\omega) &= \wp \int_0^\infty\!\!\mathrm{d}k\,\frac{g^2(k)}{k^2-\omega^2}, \\
\Gamma(\omega) &= \frac{\pi}{2}\,\frac{g^2(k)}{|\omega|}.
\end{align}
\end{subequations}
Finally, using the crossing symmetry of the propagators \ref{CrossRelation}, we get
 
\begin{equation} \label{AlphaTDep}
\alpha^{(2)}(\omega) = \frac{4mA}{(4m^2-\omega^2)\coth\left(\frac{\beta m}{2}\right)-4m\left[\Delta(\omega)+i\,\mathrm{sign}(\omega)\,\Gamma(\omega)\right]}.
\end{equation}

The last step is to use the correspondence formula \eqref{correspondence} to find a real physical atom polarizability. To do so one has to find a ratio of the partition functions ${\cal Z}/{\widehat {\cal Z}}$. In the lowest order of perturbation it is given by the formula \eqref{ZRatio2Ord}
\begin{align} 
\frac{\cal Z}{\widehat{\cal Z}} \approx \frac{\tanh({\beta m})}{\tanh\left(\frac{\beta m}{2}\right)}\left[1+\left(1-\frac{\tanh({\beta m})}{\tanh\left(\frac{\beta m}{2}\right)}\right)\delta^{(2)}\right].
\end{align}

This all means that in the lowest order of perturbation theory the Fourier transform of the linear polarizability of a two-level atom has the form
\begin{equation}\label{PolarizabilityEND}
\mathfrak{a}^{(2)}(\omega) = \frac{4mA_\mathtt{T}}{4m^2-\omega^2-4m\left[\Delta_\mathtt{T}(\omega)+i\,\mathrm{sign}(\omega)\,\Gamma_\mathtt{T}(\omega)\right]},
\end{equation}
where we introduced the effective temperature-dependent amplitude of the polarizability, as well as an effective shift and width of the resonance
\begin{subequations}
\begin{align}
A_\mathtt{T}&=\tanh({\beta m})\left[1+\left(1-\frac{\tanh({\beta m})}{\tanh\left(\frac{\beta m}{2}\right)}\right)\delta^{(2)}\right], \\
\Delta_\mathtt{T}(\omega) &=\Delta(\omega)\,\tanh\left(\frac{\beta m}{2}\right),\\
\Gamma_\mathtt{T}(\omega) &=\Gamma(\omega)\,\tanh\left(\frac{\beta m}{2}\right). \label{GammaT}
\end{align}
\end{subequations}
It is very simple now to verify that in the low temperature limit ($\beta \rightarrow \infty$) we reconstruct the polarizability at zero temperature given by us before in \cite{BiTs}. Moreover, we find that our result predicts a small correction to the behavior of the amplitude of the linear polarizability, well known from the theory of the magnetic resonance \cite{Sl90}. This correction is a second order outcome of the interaction and in the limit when $g(k)\rightarrow 0$ we reconstruct the previous result.

All physical properties of the linear polarizability of an atom at finite temperature directly follow from the equation \eqref{PolarizabilityEND}. As it is seen the magnitudes of the resonance shift and the width and the amplitude of the polarizability decrease when the temperature grows. It means that the capabilities of keeping the qubit under control at high temperature are extremely poor since the reaction of the system to the external electromagnetic field is weak (small amplitude), and the resonance is narrow (small resonance width).

\section{Thermal line narrowing}
Notice, that temperature dependence of the relative shift $\Delta_\mathtt{T}/\Delta$ and the relative width $\Gamma_\mathtt{T}/\Gamma$ of the resonance in the lowest order of perturbation do not depend on the function $g(k)$ but they depend only on the energy gap between the qubit states $\Delta E=2m$ (Fig. 1). It means that it is an universal property of physical qubits -- it does not depend on the details of experimental qubit realizations encoded in the coupling function $g(k)$. Moreover, the interaction between qubits and the quantized electromagnetic field is very similar to the interaction with the quantized vibrations of the crystal lattice (phonons). Therefore, one can imagine that the dependence on temperature in such a case would be very similar. This all means that relative temperature dependence of the shift and width of the resonance is universal for all qubits with freezed spatial degrees of freedom.

From the eq. \eqref{GammaT} one can find the sensitivity of the life-time of the exited state (the inverse of $\Gamma$) to the temperature for different experimental realizations of qubits. We define the slowing-down temperature $T_S$ -- the temperature when the life-time of the exited state becomes one order of magnitude larger than at zero temperature. The comparision of the slowing-down temperature for different realizations is given in Fig. 2.

\begin{figure}
\psfrag{T1}{$T^{(1)}_S$}
\psfrag{T2}{$T^{(2)}_S$}
\psfrag{X}{$T$}
\psfrag{Y}{$\frac{\Gamma_\mathtt{T}}{\Gamma},\frac{\Delta_\mathtt{T}}{\Delta}$}
\includegraphics[scale=0.6]{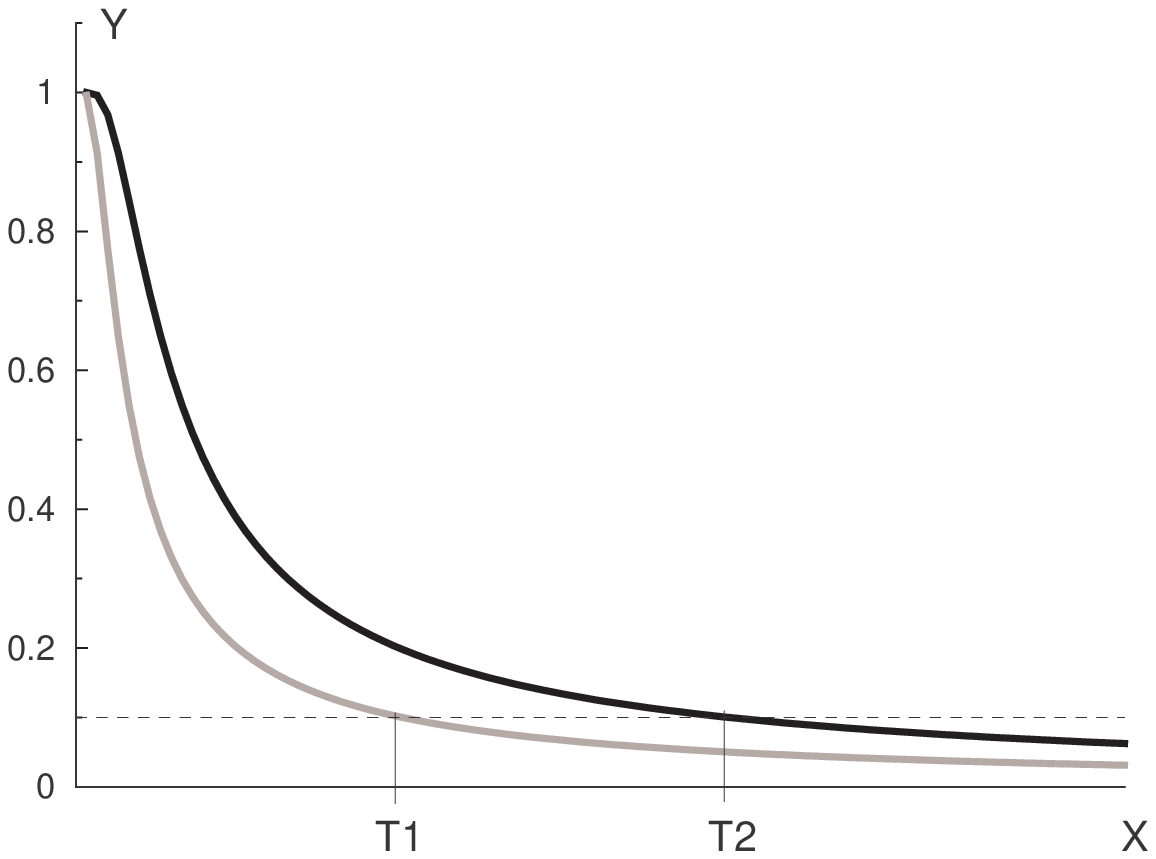}
\caption{Temperature dependence of the relative shift and relative width of the resonance for two qubits with different energy gap. Energy gap $\Delta E^{(1)}$ for the first qubit (grey line) is smaller than energy gap $\Delta E^{(2)}$ for the second qubit (black line). Therefore, the slowing-down temperature is smaller in the first case than in the second case ($T^{(1)}_S<T^{(2)}_S$).}
\end{figure}

\begin{figure}
\begin{tabular}{||c|r|c||}
\hline \hline
Qubit realization & $\Delta E=2m$ & $T_{S}$ \\
\hline \hline
Electron in quantum dot & $\sim 50\,\mathrm{meV}$ & $\sim 1500\,\mathrm{K}$ \\
$^2\mathrm{S}_{1/2}$ splitting in $^{199}\mathrm{Hg}^+$ & $167\,\mathrm{\mu eV}$ & $\sim 4.8\,\mathrm{K}$ \\
$^2\,\mathrm{S}_{1/2}$ splitting in $^{113}\mathrm{Cd}^+$ & $63\,\mathrm{\mu eV}$ & $\sim  1.8\,\mathrm{K}$ \\
$^2\mathrm{S}_{1/2}$ splitting in $^{87}\mathrm{Rb}^+$ & $28\,\mathrm{\mu eV}$ & $\sim 0.8\,\mathrm{K}$ \\
\hline \hline
\end{tabular}
\caption{Comparison of the slowing-down temperature $T_S\approx 2.5\times \frac{\Delta E}{k}$ (temperature when life-time of exiting state becomes one order of magnitude larger than in zero temperature) for different physical realizations of qubits \cite{Ha03,El04,Ma73,Ta96,Es61,Pe62}.} \label{tabelka}
\end{figure}

In conclusion, let us concentrate on the problem of the width of the resonance. As was shown above, the resonance width unexpectedly decreases when the temperature grows. Such a behavior is contrary to the natural expectations. In typical situations we expect broadening of the resonance rather than narrowing. The best known examples where the resonance broads with temperature are obviously heated gases where, according to the Doppler effect and the chaotic motion of the molecules, all spectral lines become wider when the temperature grows.

In the literature there exist some theoretical predictions on the narrowing of the resonance known as the motional narrowing phenomenon. First theoretical analysis of the motional narrowing was given in the fundamental paper on nuclear resonance by Bloembergen {\it et al.} \cite{Bl48}. The narrowing of the resonance is explained due to the influence of the thermal motion on the strength of the spin-spin interactions. It turns out that the thermal motion effectively leads to decreasing of the damping effects induced by such interactions. The general problem of the motional narrowing was deeply studied and generalized in the language of the stochastic processes under the assumption that the system is subjected to a totally chaotic fluctuating environment \cite{Ku62}. There exist lot of physical systems where the motional narrowing appears \cite{Dy71,Ox79,Eb84,Bert06}.

The origin of the narrowing of the resonance predicted in presented paper is completely different. Notice that in the studied situation there is only one freezed qubit. Therefore there is no place for any influence of the Doppler effect or of any interaction with other qubits. This suggest that it should be called motionless narrowing rather than motional one, since it is clear that freezing of spatial degrees of freedom is a main source of the narrowing of the resonance. 

\begin{acknowledgments}
I thank professor Iwo Bialynicki-Birula for his incisive questions, useful comments, and fruitful discussion. This work was supported by a grant from the Polish Ministry of Science and Higher Education.
\end{acknowledgments}

\appendix

\section{Relations between partition functions}
In this appendix we derive a direct relations between the partition functions ${\cal Z}$ and $\widehat{\cal Z}$. These relations are exact. They hold in any order of perturbation theory.
\subsection{Difference of partition functions}
Let us consider a family of the systems parametrized by $\lambda$ described by the following statistical Hamiltonian
\begin{subequations}
\begin{align}
\widehat{\cal K}_\lambda = \widehat{\cal K}_0 + \lambda \widehat{\cal H}_\mathtt{I}
\end{align}
and assume that each of those systems is in the quantum state described by a density matrix
\begin{align}
\widehat{\rho}_\lambda=\mathrm{e}^{-\beta \widehat{\cal K}_\lambda}, \qquad \widehat{\cal Z}_\lambda = \mathrm{tr}\left[\mathrm{e}^{-\beta \widehat{\cal K}_\lambda}\right].
\end{align}
\end{subequations}
It is clear that for $\lambda=1$ we recover the situation described by the formulas \eqref{Stat1} and for $\lambda=0$ by the formulas \eqref{Stat9}. Let us also introduce the second-quantized counterpart of this family
\begin{align}
{\cal K}_\lambda &= {\cal K}_0 + \lambda {\cal H}_\mathtt{I}, \\
\rho_\lambda &=\mathrm{e}^{-\beta {\cal K}_\lambda}, \qquad {\cal Z}_\lambda = \mathrm{tr}\left[\mathrm{e}^{-\beta {\cal K}_\lambda}\right].
\end{align}
One can easily get a derivative of the partition function with respect to $\lambda$ for these two families
\begin{subequations}
\begin{align}
\frac{\partial {\cal Z}_\lambda}{\partial \lambda}&=-\beta \,\mathrm{tr}\left[\mathrm{e}^{-\beta {\cal K}_\lambda}{\cal H}_\mathtt{I}\right]=-\frac{\beta}{\lambda}{\cal Z}_\lambda \, \langle \lambda {\cal H}_\mathtt{I}\rangle_\lambda, \\
\frac{\partial \widehat{\cal Z}_\lambda}{\partial \lambda}&=-\beta \,\mathrm{tr}\left[\mathrm{e}^{-\beta \widehat{\cal K}_\lambda}\widehat{\cal H}_\mathtt{I}\right]=-\frac{\beta}{\lambda}\widehat{\cal Z}_\lambda\, \langle\!\langle \lambda \widehat{\cal H}_\mathtt{I}\rangle\!\rangle_\lambda.
\end{align}
\end{subequations}
The argument leading to the formula \eqref{correspondence} is also valid for the systems with any $\lambda$. Since the interaction Hamiltonian ${\cal H}_\mathtt{I}$ has the needed property (in the qubit subspace it is represented by a traceless matrix proportional to the Pauli $\sigma_x$ matrix), one has the important relation
\begin{align}
\langle\!\langle \lambda \widehat{\cal H}_\mathtt{I}\rangle\!\rangle_\lambda =
\frac{{\cal Z}_\lambda}{{\widehat{\cal Z}}_\lambda}\langle \lambda {\cal H}_\mathtt{I}\rangle_\lambda.
\end{align}
This immediately leads us to the nontrivial equality of the derivatives
\begin{align}
\frac{\partial {\cal Z}_\lambda}{\partial \lambda}=\frac{\partial \widehat{\cal Z}_\lambda}{\partial \lambda}.
\end{align}
Therefore one gets a crucial relation
\begin{align} \label{PartitionFDifference}
{\cal Z}-{\cal Z}_0 = \widehat{\cal Z}-\widehat{\cal Z}_0.
\end{align}
It means that the difference between the partition functions ${\cal Z}$ and $\widehat{\cal Z}$ is the same as a difference between the partition functions for noninteracting systems  ${\cal Z}_0$ and $\widehat{\cal Z}_0$, which can be calculated directly.

\subsection{Ratio of partition functions}
From the general rules of the thermal quantum field theory we know that the partition function ${\cal Z}$ can be expressed as a sum of all vacum-like Feynman diagrams
\begin{subequations}
\begin{align}
{\cal Z}={\cal Z}_0\left(1+ \delta\right),
\end{align}
where
\begin{align}
-\delta = \includegraphics[scale=0.7]{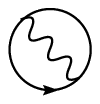}+\includegraphics[scale=0.7]{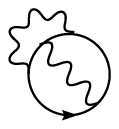}+\includegraphics[scale=0.7]{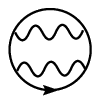}+\includegraphics[scale=0.7]{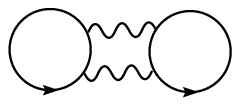}\ldots
\end{align}
\end{subequations}
With the equation \eqref{PartitionFDifference} one can find that
\begin{equation}
\widehat{\cal Z} = \widehat{\cal Z}_0+{\cal Z}-{\cal Z}_0 = \widehat{\cal Z}_0+{\cal Z}_0\,\delta
\end{equation}
and therefore the ratio of the partition functions is given by
\begin{equation} \label{ZRatioNotExp}
\frac{\cal Z}{\widehat{\cal Z}} = \frac{{\cal Z}_0}{\widehat{\cal Z}_0}\,\frac{1 + \delta}{1+\frac{{\cal Z}_0}{\widehat{\cal Z}_0}\delta}.
\end{equation}
This result is correct in any order of perturbation theory and it is easy to verify that the ratio ${\cal Z}_0/\widehat{\cal Z}_0$ of the partition functions for noniteracting systems reads
\begin{align}
\frac{{\cal Z}_0}{\widehat{\cal Z}_0}=\frac{\mathrm{e}^{-\beta m}+\mathrm{e}^{\beta m}+2}{\mathrm{e}^{-\beta m}+\mathrm{e}^{\beta m}}=\frac{\tanh({\beta m})}{\tanh\left(\frac{\beta m}{2}\right)}.
\end{align}
To find an appropriate formula in a given order of perturbation one should expand the relation \eqref{ZRatioNotExp} in powers of the coupling constant (the formfactor $g(k)$) and truncate the series at the appropriate position. For example, in the lowest (second) order one finds
\begin{align} \label{ZRatio2Ord}
\frac{\cal Z}{\widehat{\cal Z}} \approx \frac{{\cal Z}_0}{\widehat{\cal Z}_0}\left[1+\left(1-\frac{{\cal Z}_0}{\widehat{\cal Z}_0}\right)\delta^{(2)}\right],
\end{align}
where $\delta^{(2)}$ is represented only by one Feynman diagram \eqref{TempDeltaOmegaA} and it is given by \eqref{DeltaResult2} evaluated in the Appendix B.

\section{Second order calculations}
\subsection{Photon self-energy function}
The temperature self-energy function of the photon in the lowest order of perturbation theory is represented by the following Feynman diagram
\begin{subequations}
\begin{align} \label{TempPOmegaA}
-\widetilde{\mathrm{P}}^{(2)}(\omega_n) &= \includegraphics[scale=0.8]{pi00A.eps} \\[10pt]
&=-\frac{1}{\beta}\sum_{n'} \mathrm{Tr}\big\{
\sigma_x \widetilde{S}(\omega_{n'}+\omega_n)\sigma_x \widetilde{S}(\omega_{n'})\big\}.
\end{align}
Depending on the quantum statistics (see eq. \eqref{FT10}), the frequencies $\omega_n$ and $\omega_{n'}$ are
\begin{equation} \label{TempPOmega}
\omega_n = \frac{2 n \pi}{\beta}, \qquad \omega_{n'} = \frac{(2n'+1) \pi}{\beta}.
\end{equation}
\end{subequations}
One can easily transform the relation \eqref{TempPOmegaA} to a simpler form
\begin{multline} \label{P2BeforeSum}
\widetilde{\mathrm{P}}^{(2)}(\omega_n) =  \frac{1}{\beta} \sum_{n'=-\infty}^\infty \left[\frac{1}{i\omega_{n'}+i\omega_n+m}\frac{1}{i\omega_{n'}-m} \right. \\
\left.+\frac{1}{i\omega_{n'} +i\omega_n -m}\frac{1}{i\omega_{n'} +m}\right].
\end{multline}
The summation over $n'$ can be done with the exploitation of the following mathematical identity
\begin{multline}  \label{SumFermion}
\sum_{n'} \frac{1}{i\omega_{n'}-a}\,\frac{1}{i\omega_{n'}-b}= -\frac{\beta}{2}\frac{\tanh\left(\frac{\beta b}{2}\right)-\tanh\left(\frac{\beta a}{2}\right)}{b-a}
\end{multline}
which is valid for $\omega_{n'}=(2 n' +1)\pi/\beta$.

The temperature self-energy function of the photon in the lowest order of perturbation theory has the form
\begin{multline}
\widetilde{\mathrm{P}}^{(2)}(\omega_n) 
=-\frac{1}{2}\left[\frac{\tanh\left(\frac{\beta m}{2}\right)+\tanh\left(\frac{\beta}{2}(m+i\omega_n)\right)}{i\omega_n+2m} \right. \\
\left. -\frac{\tanh\left(\frac{\beta m}{2}\right)+\tanh\left(\frac{\beta}{2}(m-i\omega_n)\right)}{i\omega_n-2m}\right].
\end{multline}
Conclusively, according to the periodicity of the hyperbolic function and the relation \eqref{TempPOmega}, the temperature self-energy function has the form
\begin{align} \label{TLA-P-2TApp}
\widetilde{\mathrm{P}}^{(2)}(\omega_n)&=- \frac{4m}{4m^2+\omega_n^2}\tanh\left(\frac{\beta m}{2}\right).
\end{align}
\subsection{Correction to the partition function}
The first correction to the partition function of the interacting system is represented by the following Feynman diagram
\begin{subequations}
 
\begin{multline} \label{TempDeltaOmegaA}
-\delta^{(2)} = \includegraphics[scale=0.8]{vac1.eps} \\
=\frac{1}{\beta^2}\sum_{n}\sum_{n'}\int_0^\infty \!\!\mathrm{d}k\!\int_0^\infty \!\!\mathrm{d}k'\,\widetilde{D}(k,k',\omega_n) \\ \times\mathrm{tr}\left\{V(k)\widetilde{S}(\omega_{n'}+\omega_n)V(k')\widetilde{S}(\omega_{n'})\right\}.
\end{multline}
Summations run over the boson-like $\omega_n$ and the fermion-like $\omega_{n'}$ 
\begin{equation} \label{TempDeltaOmegaB}
\omega_n = \frac{2 n \pi}{\beta}, \qquad \omega_{n'} = \frac{(2 n'+1) \pi}{\beta}.
\end{equation}
\end{subequations}
However, the summation over $n'$ can be easily expressed by the photon self-energy function in the second order of perturbation theory \eqref{TempPOmegaA} evaluated before
\begin{align}
-\delta^{(2)} &= -\frac{1}{\beta}\int_0^\infty \!\!\mathrm{d}k\,g^2(k)\,\sum_{n} \frac{\widetilde{\mathrm{P}}^{(2)}(\omega_n)}{\omega_n^2+k^2} \nonumber \\ &=\frac{4m}{\beta}\mathrm{tanh}\left(\frac{\beta m}{2}\right) \int_0^\infty \!\!\mathrm{d}k\,g^2(k)\nonumber \\
&\qquad\qquad\qquad\times\sum_{n} \frac{1}{\omega_n^2+4m^2}\,\frac{1}{\omega_n^2+k^2}.
\end{align}
The summation over $n$ one can do similarly as in formula \eqref{P2BeforeSum}. For the bosonic frequencies $(\omega_{n}=2 n\pi/\beta)$ the appropriate summation rule reads 
\begin{subequations} \label{SumBoson}
\begin{equation} 
\sum_{n} \frac{1}{i\omega_{n}-a}\,\frac{1}{i\omega_{n}-b}=-\frac{\beta}{2}\frac{\coth\left(\frac{\beta b}{2}\right)-\coth\left(\frac{\beta a}{2}\right)}{b-a}
\end{equation}
\end{subequations}
After performing the appropriate summations one gets the final expression for the first correction to the partition function
\begin{multline} \label{DeltaResult2}
\delta^{(2)} = \mathrm{tanh}\left(\frac{\beta m}{2}\right)\int_0^\infty \!\!\!\mathrm{d}k\,\,\frac{g^2(k)}{k(k^2-4m^2)} \\
\times\left[2m\coth\left(\frac{\beta k}{2}\right)-k \coth\left(\beta m\right)\right].
\end{multline}
 
At zero temperature limit $(\beta \rightarrow \infty)$, this correction is finite and negative. It is a simple exercise to find that in this limit it has a form
\begin{align}
\lim_{\beta\rightarrow \infty}\delta^{(2)} &= -\int_0^\infty \!\!\!\mathrm{d}k\,\,\frac{g^2(k)}{k(2m+k)}.
\end{align}
It is worthwhile to notice that even near the resonance, when $k\approx 2m$, the integral \eqref{DeltaResult2} is well defined, because the dangerous pole in the denominator is canceled by an appropriate term in the numerator of the expression.

\end{document}